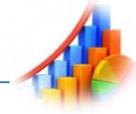

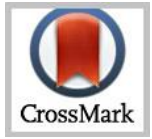

# Deployment of Advanced and Intelligent Logistics Vehicles with Enhanced Tracking and Security Features

## Iqtiar Md Siddique[1*], Selim Molla[2], MD Rakib Hasan[3], Anamika Ahmed Siddique[1]

[1]Department of Mechanical and Aerospace Engineering, University of Texas, El Paso, US
[2]Department of Computational Science, University of Texas, El Paso, US
[3]Department of Mechanical Engineering, Lamar University, US

[*]Corresponding Author's Email: iqtiar.siddique@gmail.com

**ABSTRACT:** This study focuses on the implementation of modern and intelligent logistics vehicles equipped with advanced tracking and security features. In response to the evolving landscape of logistics management, the proposed system integrates cutting-edge technologies to enhance efficiency and ensure the security of the entire logistics process. The core component of this implementation is the incorporation of state-of-the-art tracking mechanisms, enabling real-time monitoring of vehicle locations and movements. Furthermore, the system addresses the paramount concern of security by introducing advanced security measures. Through the utilization of sophisticated tracking technologies and security protocols, the proposed logistics vehicles aim to safeguard both customer and provider data. The implementation includes the integration of QR code concepts, creating a binary image system that conceals sensitive information and ensures access only to authorized users. In addition to tracking and security, the study delves into the realm of information mining, employing techniques such as classification, clustering, and recommendation to extract meaningful patterns from vast datasets. Collaborative filtering techniques are incorporated to enhance customer experience by recommending services based on user preferences and historical data. This abstract encapsulates the comprehensive approach of deploying modern logistics vehicles, emphasizing their intelligence through advanced tracking, robust security measures, and data-driven insights. The proposed system aims to revolutionize logistics management, providing a seamless and secure experience for both customers and service providers in the dynamic logistics landscape.

## 1. INTRODUCTION

Logistics plays a crucial role in managing the efficient flow of materials, work-in-process, and finished inventories at the lowest total cost. It involves coordinating various elements such as order processing, inventory management, transportation, warehousing, materials handling, and packaging across a network of facilities. Logistics data management arrangements include various modules such as system management, resource management, customer management, contract management, exception management, manufacturing management, business management, and billing management. These subsystems operate with distinct functionalities, and information systems act as the cohesive thread that integrates logistics practices into a unified process. In the proposed system, the primary focus is on ensuring end-to-end security for customer and provider data using the QR code concept. This involves employing a binary image system within the QR code to conceal both customer and provider information, accessible only to authorized users. Customer data mining is facilitated through collaborative filtering, where the recommendation of vehicles is based on provider preferences. The central principle is to propose vehicles according to provider benefits, enabling the system to identify customer interests and recommend relevant services. Additionally, the system incorporates spam service provider detection to enhance the quality of recommendations. Information mining, a process of extracting meaningful patterns and relationships from vast datasets, is employed in the system to predict future trends. Various techniques, including classification, clustering, and recommendation, are integrated into the information mining process. Collaborative filtering, a technique used in recommender systems, relies on users' active engagement, an effective method to address users' interests, and algorithms that match individuals with similar interests. The Euclidean remoteness, computing the straight-line distance relating two viewpoints, is utilized for similarity calculations in collaborative filtering.

Furthermore, the system incorporates the use of stop words, commonly used words ignored by search engines during indexing and retrieval. These stop words are excluded to enhance search efficiency and are particularly relevant in natural language processing (NLP). The inclusion of Quick Response (QR) codes, a type of 2D barcode, enhances accessibility to information through smartphones.

In logistic systems, there is a growing focus on public transportation services, and two main categories are identified: those responding to real-time requests and those based on historical mobility patterns using GPS data. The overarching goal of the anticipated system is to prepare optimal logistics services towards customers at minimal costs while enabling real-time tracking of vehicle locations. The utilization of QR codes, collaborative filtering, and information mining enhances the overall efficiency and security of the logistics management system.

## 2. LITERATURE SURVEY

The Vehicle Tracking System is a device commonly employing GPS technology, affixed to a vehicle to monitor its location. Initially, it captures satellite signals to determine its precise coordinates in terms of latitude and longitude. These coordinates are typically visualized on a computer screen, and mapping software aids in displaying the exact position of the vehicle. This technology enables users to access information about a vehicle, including its position, speed, distance traveled, and duration of each stoppage. The central operating center facilitates this access by users, who can input their mobile number through mobile phones or websites using SMS or the Internet. Vehicle tracking technology proves beneficial for overseeing both commercial and passenger vehicles. In the context of personal vehicle tracking, it serves the purpose of locating a stolen vehicle with pinpoint accuracy, providing the exact location. An effective framework that analyzes driver and traveler preferences using a transformative game approach (Qiao et al., 2015). This approach aims to enhance drivers' earnings and reduce travelers' costs through an efficient dispatch model. The model is capable of significantly reducing the time spent locating travelers and increasing driver profits by at least 18%, consequently minimizing travelers' waiting times. On the Online to Offline (O2O) taxi business, addressing the conflicting interests of travelers, taxi drivers, and the organization (Tian et al., 2013). For both sharing and non-sharing taxi dispatches, the paper introduces stable marriage approaches to manage unequal quantities of traveler requests and taxis. The proposed algorithms demonstrate significant improvements in taxi satisfaction, as compared to existing methods. The evolving nature of the mobility industry, predicting significant changes and advancements (James & Lam, 2017). The authors argue that city development will replace regional considerations as the key factor influencing mobility behavior, requiring industry players to engage in multifaceted strategies. R. A. Vasco, R. Morabito et al. [4] propose a system to address the Dynamic Vehicle Allocation Problem (DVAP) in road transportation, specifically for full truckloads between terminals (Vasco & Morabito, 2016). The system involves multi-period resource assignment, considering the movements of a fleet of vehicles across geographically dispersed terminals. The proposed algorithms, including GRASP and simulated annealing meta-heuristics, prove effective in making timely decisions for a transportation company in Brazil. Revisit the taxi dispatch issue in the Online to Offline (O2O) taxi business, considering the conflicting interests of travelers, taxi drivers, and the organization (Zheng & Wu, 2017). The proposed algorithms, both for sharing and non-sharing taxi dispatches, demonstrate improved performance in terms of dispatch delay and traveler satisfaction. An extensive review of the vehicle routing problem, considering practical applications in various sectors over the past 50 years (Coelho et al., 2016). The paper discusses the use of operations research methods to optimize vehicle routing, leading to significant cost savings in sectors such as oil, gas, retail, waste collection, and food delivery. The challenge of assigning stock keeping units to different types of distribution centers in the retail industry (Holzapfel et al., 2018). The problem is approached through a Mixed Integer Programming (MIP) solution, considering the interdependencies between inbound and outbound transportation, inshore logistics, inventory, and picking costs. The proposed method, operated in an actual example in a European grocery marketing group, develops substantial fee funds of 6% in absolute working costs. The approach provides insights into design structures and related issues through in-depth analyses of results and sensitivity assessments. Delicately articulate insights across four distinct papers concerning manufacturing excellence, scheduling operations, and equipment efficiency. These contributions are valuable considerations for overall equipment selections in the development of an intelligent logistics system, with vehicle tracking identified as a crucial requirement (Ullah et al., 2023; Ullah et al., 2023; Ullah et al., 2023). Addresses the significance of cryptocurrency systems, particularly in the electronics sector, influencing the selection of security materials for vehicle security systems (Rahman et al., 2023). Emphasize environmental factors and safety risk assessments in the context of human and environmental considerations, which significantly impact the considerations for a vehicle tracking system (Fayshal et al., 2023; Molla et al., 2024). Evidence of RFID technology for warehouse management through an android application, demonstrating its profound impact on the electronics industry, especially in the context of intelligent system tracking (Kamal et al., 2019). An informative process flow chart for a jute mill, enriching industry data and aiding in vehicle tracking research (Shakil et al., 2013). Discuss electricity generation from moving vehicles, proposing its potential application for machine continuity in a factory, aligning with the objectives of tracking and intelligent logistics management (Hossain et al., 2023). The importance of medical textiles with plantable and implantable options, serving as a focal point for future research (Siddique et al., 2023). Employs remote sensing methods in land surface interpretation, contributing to intelligent logistics time tracking systems, particularly with remote sensing devices (Mustaquim, 2024). In a comprehensive discussion on ergonomics factors, influencing worker efficiency in electronics plants and, by extension, impacting vehicle tracking systems, where visualization systems have specific requirements (Parvez et al., 2022; Parvez et al., 2022). Explores the impact of supplier selection on the electronics sector, recognizing its significant role in product procurement for logistics and intelligent tracking systems (Khofiyah et al., 2021). Implement machine learning algorithms, particularly focusing on predicting performance in the production operations sector, especially in scenarios involving substantial amounts of big data. This study is deemed highly beneficial, laying a foundation for potential expansions in future research endeavors (Rahman et al., 2023; Rahman et al., 2023). Provide substantial insights into COVID data in the United States and globally, offering valuable information for adhering to COVID protocols in the development and maintenance of rules and regulations in any intelligent vehicle system (Molla et al., 2023).

## 3. PROPOSED WORK

To address the constraints of traditional transportation strategic frameworks, a proposed online solution aims to enable both clients and assistance providers to roadway automobiles during conveyance. The system also strives to offer optimal services to clients at the lowest cost by recommending available service providers at preferred rates. In essence, the theoretical framework involves the dynamic allocation of customer requests and responses from service providers, tracking the logistics of the vehicle system, and providing evidence in the model of QR codes. The outlined work primarily consists of four modules: admin, customer, driver, and service provider. Each module serves specific functions within the system. The administrative module oversees the overall system, while the customer module allows clients to interact with the system, track vehicles, and receive services at competitive costs. The driver module facilitates the participation of drivers in the transportation process, and the service provider module handles the allocation and response to customer requests. This proposed solution strives to enhance the efficiency of transportation logistics by incorporating real-time tracking and cost-effective recommendations. The modular structure ensures a well-organized system that caters to the requests of clients and facility contributors while optimizing the overall transportation experience.

Admin: Within this system, the administrator is tasked with granting authentication permissions to service providers and possesses the capability to observe information pertaining to vehicles, customers, providers, detect spam service providers, and access the ranking of service providers.

Service Provider: Within this framework, service providers have the authority to add vehicles and drivers. They can also view customer requests, dispatch notifications to drivers, and access information on scheduled vehicles and historical data.

Customer: In this setup, customers could peruse available vehicles, search for specific ones, request a vehicle, track its location on a map, and make payments to service providers. Additionally, customers can contribute reviews to the system and exchange data in the shape of QR codes.

Driver: Within this arrangement, drivers be capable of reviewing incoming requests and accessing the schedule for assigned vehicles.

## 4. SYSTEM ARCHITECTURE

In the existing strategic administration framework, clients are faced with the challenge of searching for suppliers and suitable transportation means, leading to increased waiting times and a lack of real-time tracking for transported materials. The primary focus of our technique is to ensure end-to-end security for both customer and provider data through the implementation of QR code technology. The QR code, in parallel images, conceals customer and provider information, with access granted only to authorized customers. To enhance customer interest mining, we employ a collaborative filtering method, which revolves around recommending vehicles based on provider preferences. This recommendation process aids in identifying customer interests and providing relevant suggestions. Consumer assistance, a term used in the context of interest drilling, involves offering guidance or solutions to issues. Recommendations play a pivotal role, serving as a client interest indicator, guiding new customers to utilize professional service vehicles. The paramount concern is to maintain the security of customer and provider data throughout the entire process using the QR code concept. The proposed system comprises four main modules: admin, service provider, customer, and driver. The admin module is responsible for providing authentication permissions to providers, along with the ability to view vehicles, customers, providers, detect spam service providers, and rank service providers. A noteworthy project on data retrieval approaches, with coding technology pivotal for our research (Noman et al., 2020; Noman et al., 2018). Employ value stream mapping with a robust mathematical process, offering utility in research, especially in scenarios where extensive vehicle production is essential for the tracking system as given in Figure 1 (Ullah et al., 2024).

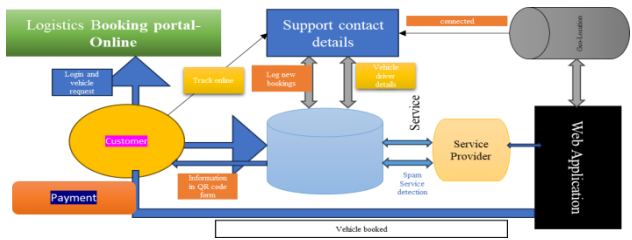

***Figure 1:*** *System Architectural View of Proposed Framework.*

The provider module allows providers to add automobiles and motorists, view customer requests, and send notifications to drivers. Providers can also access scheduled vehicles and view historical data. Within the customer module, clients can view and search for vehicles, request transportation, track vehicles on a map, make payments to service providers, provide system reviews, and exchange information using QR codes. The driver module enables drivers to view requests, schedule vehicles, and participate in the overall transportation process. This comprehensive system aims to streamline the interaction between clients, providers, and drivers while ensuring the security and efficiency of the entire process through QR code implementation. The advantage of the proposed system is to provide the best possible logistics services to the customer at lowest cost. Allowing the customer to trace the current location of vehicle on the map. Implementing a QR code-based system to ensure comprehensive security for both customer and provider data. Additionally, suggesting the optimal service provider in proximity based on user preferences. This strategy aims to safeguard information from end to end, offering a secure and personalized experience for users.

## 5. EXPRIMENTAL SETUP

- System Description: By Installing and rectifying the transmitter within the vehicle tracking system by employing a push-pull button mechanism for device activation and deactivation. Subsequently, in the second step, establish a connection between the receiver and a computer via an RS232 cable, initiating the receiver using a push-pull button in conjunction with the computer's startup. Moving on to the third step, the transmitter incorporates a GSM modem to relay the vehicle's location remotely to a designated mobile device, presenting the owner with a message containing latitude and longitude coordinates. Simultaneously, in the fourth step, this information is transmitted to the computer via an RS232 cable for display and output purposes. In the proposed system, a flow chart for transmitter is given at Figure 2:

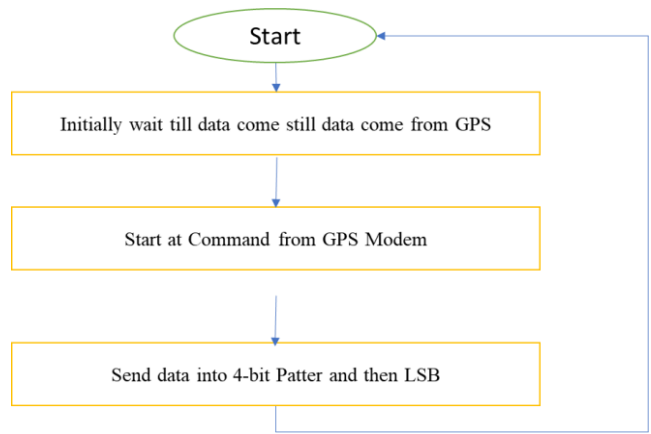

***Figure 2:*** *Transmitter Flowchart.*

- Statistical Model: Let's contemplate the entity denoted as M, representing an automated system designed to suggest vehicles to customers. S is defined as {E, J, G, e}.

- Input: Identify the inputs E= e1, e2, e3 ...., en— E represents a collection of functions designed to carry out commands.

- J= j1, j2, j3 Define the sets of inputs as the function set
- G= g1, g2, g3 Define the sets of outputs as the function set,
- p = Program completion or End.
- $M_1$= I, E, O
- I = Customer's inquiry, namely the query submitted by the customer.
- O = Generate the suggested result for the specified request, namely, the recommendation of a suitable vehicle.
- E = Functions are employed to obtain the output, specifically through collaborative filtering.

Mapping Diagram:

Mapping diagram is given in Figure 3.

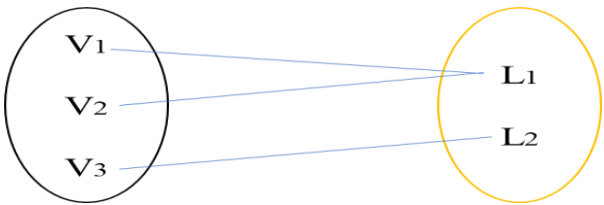

*Figure 3: Mapping Diagram.*

Where, Users as V

L=location query.

V1= Construct a query related to the correct geographical position (L1)

V2= Right Location Query (L1)

V3= Construct a query related to the Wrong geographical position (L2)

Establish a set notation as follows: M = {a, e, X, Y}.

In the given context, 'a' represents the initial point of the program.

1. Access the webpage by logging in.
2. View vehicle recommendation as per location, track vehicle location. Do payment and give ratings for the service. e = End of the program.

Retrieve comprehensive information regarding vehicles from the service provider. Users have the capability to access and review details about both drivers and vehicles upon booking. The system suggests vehicles based on cost and location.

Success Conditions:

Efficiently search for the necessary information within the database.

Users experience prompt results aligned with their specific requirements.

Failure Conditions:

Extensive databases may result in prolonged information retrieval times.

Potential hardware failures.

Occurrence of software malfunctions.

Formulate a mathematical representation in the form of an equation for a Logistic System and rephrase it to ensure originality:-

Given: Xq= Determine whether a request "q" belonging to the set "Q" will result in the allocation of a vehicle or not.

Tr= Required Time for Truck or vehicle request,

Cv=Vehicle types of cost.

Dr= Length of the route.

Ct=Expenditure for Trip.

Fu=Fuel Used

Equation: -If $X_q$ is accepted with Tr then Ct is calculated by following equation: Ct=Dr* Cv

Once the trip cost has been computed, determine the fuel expenses by employing the subsequent equations: Fu=$C_t$*$D_r$

In our experimental setup, as detailed in Table 1, we determine the number of reviews from various users. In this configuration, there are 100 positive reviews and 60 negative reviews, representing a twofold increase compared to the original values.

| Sr. No | Positive Review | Negative Review |
|---|---|---|
| 1 | 50 | 30 |

## 6. DATASET

In the proposed system, we need a data set. In the proposed system, we need a data set. of keyword of positive and negative keyword for classification of review.

*Table 1: Dataset of Positive and Negative Keywords.*

| S. No. | Positive Keyword | Negative Keyword |
|---|---|---|
| 1 | Good | Sad |
| 2 | Great | Bad |
| 3 | Happy | Poor |
| 4 | Active | Useless |
| 5 | Nice | Cold |
| 6 | Believe | Cry |

 

## 7. RESULT

From the above data, as shown in Table 1, we find out number of reviews of different user. Here 50 positive reviews and 30 negative reviews are shown in below Figure 4:

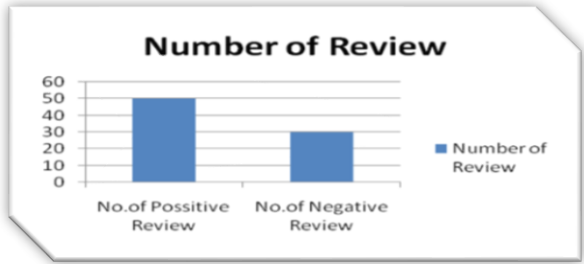

***Figure 4:*** *Number of Positive and Negative Reviews.*

## 8. LIMITATIONS

Despite the promising prospects of deploying advanced and intelligent logistics vehicles with enhanced tracking and security features, several limitations need consideration. Firstly, the high initial costs associated with integrating sophisticated technologies may pose a significant financial barrier for some logistics companies, limiting widespread adoption. Secondly, the reliance on complex electronic systems makes these vehicles susceptible to malfunctions or technical glitches, potentially leading to service disruptions and increased maintenance costs. Additionally, concerns about data privacy and security may arise due to the extensive collection and transmission of sensitive information, necessitating robust cybersecurity measures. Geographical constraints and varying infrastructure development levels across regions may also limit the effectiveness of advanced tracking features in certain areas. Finally, regulatory frameworks and standardization issues must be addressed to ensure seamless interoperability and compliance, presenting a challenge in a landscape with evolving and diverse regulations.

## 9. CONCLUSION

The envisioned system comprises service providers, customers, administrators, and drivers, with the administrator playing a pivotal role. In this context, customers can reserve vehicles and track their current location through GPS monitoring. Logistics involves the coordination of techniques to govern the transfer besides spatial planning of raw supplies, WIP, and completed supplies at the most economical overall budget. The proposed system is centered around the transportation of goods, raw materials, relocation of household items, and furniture during the moving process. It also encompasses the administration of group handing out, stock, carrying, and the integration of distribution, items holding, and binding, orchestrated across a setup of capabilities. The system aims to ensure comprehensive security for customer and provider data through the implementation of QR code technology. Additionally, it recommends the closest and best service provider based on customer preferences.

## 10. FUTURE WORK

The future direction of the project involves a multifaceted approach to advance and refine logistics operations. Integration of cutting-edge AI algorithms will elevate decision-making capabilities, optimizing route planning and vehicle scheduling. Autonomous vehicle technology will be explored to enhance efficiency, reduce costs, and improve overall reliability. Blockchain implementation is envisaged for heightened security, ensuring transparent and secure tracking throughout the supply chain. Real-time monitoring through IoT devices will provide comprehensive visibility into cargo status, vehicle health, and environmental conditions. Emphasis will be placed on designing energy-efficient and sustainable logistics vehicles, incorporating alternative energy sources. Predictive maintenance solutions, advanced security measures, and telematics solutions will contribute to the continuous improvement of logistics operations. Human-machine collaboration will be a focus, fostering efficient communication between intelligent vehicles and human operators. Staying attuned to regulatory compliance and standards will be pivotal to ensure the seamless integration of advanced logistics vehicles within legal and safety frameworks.